\documentclass[10pt,superscriptaddress,twocolumn]{revtex4-1}

\usepackage{graphicx}

\usepackage{amssymb}

\usepackage{amsmath}

\begin{document}

\title{Local Detection of Quantum Correlations with a Single Trapped Ion}

\author{M. Gessner}

\email{manuel.gessner@physik.uni-freiburg.de} 

\affiliation{Department of Physics, University of California, Berkeley, California 94720, USA}

\affiliation{Physikalisches Institut, Universit\"at Freiburg, Hermann-Herder-Strasse 3, D-79104 Freiburg, Germany}

\author{M. Ramm}

\affiliation{Department of Physics, University of California, Berkeley, California 94720, USA}

\author{T. Pruttivarasin}

\affiliation{Department of Physics, University of California, Berkeley, California 94720, USA}

\author{A. Buchleitner}

\affiliation{Physikalisches Institut, Universit\"at Freiburg, Hermann-Herder-Strasse 3, D-79104 Freiburg, Germany}

\author{H.-P. Breuer}

\affiliation{Physikalisches Institut, Universit\"at Freiburg, Hermann-Herder-Strasse 3, D-79104 Freiburg, Germany}

\author{H. H\"affner}

\affiliation{Department of Physics, University of California, Berkeley, California 94720, USA}

\date{\today}                                           

\maketitle

\textbf{As one of the most striking features of quantum mechanics, quantum correlations are at the heart of quantum information science \cite{HHHH,MODIREVIEW,NIELSEN}. Detection of correlations usually requires access to all the correlated subsystems \cite{AUCCAISE,SILVA}. However, in many realistic scenarios this is not feasible since only some of the subsystems can be controlled and measured. Such cases can be treated as open quantum systems interacting with an inaccessible environment \cite{BREUERBOOK}. Initial system-environment correlations play a fundamental role for the dynamics of open quantum systems \cite{PECHUKAS, LINDBLAD, BREUERBOOK, WITNESS}.
Following a recent proposal \cite{GB,GB1}, we exploit the impact of the correlations on the open-system dynamics to detect system-environment quantum correlations without accessing the environment. We use two degrees of freedom of a trapped ion to model an open system and its environment. 
The present method does not require any assumptions about the environment, the interaction or the initial state and therefore provides a versatile tool for the study of quantum systems.}

Quantum correlations are particularly important in the context of quantum simulation \cite{CHRISTIAN,SCHINDLER,MYATT}, quantum phase transitions \cite{Fazio,Dillenschneider}, as well as for quantum computation \cite{HARTMUTREVIEW}. In these experiments, one typically strives to study quantum many-body dynamics in high-dimensional Hilbert spaces. However, it is precisely in these complex systems where it becomes increasingly difficult to experimentally detect quantum correlations since standard methods such as full state tomography are impractical \cite{ABUREVIEW}. Therefore, it seems natural to restrict oneself to measurements of a smaller controllable subsystem \cite{SINGLESPIN}. Similarly, in quantum communication protocols, each party has access only to its part of the shared correlated state but may want to confirm the presence of quantum correlations locally \cite{NIELSEN}. All these situations can be described in the framework of a well-controlled open quantum system in contact with an inaccessible environment \cite{BREUERBOOK}.

Initial system-environment correlations can significantly change the dynamics of open systems \cite{PECHUKAS, LINDBLAD, BREUERBOOK, WITNESS}. The standard master equation approach to open systems assumes an initial state with vanishing total correlations, which may not be appropriate unless a product state is explicitly prepared \cite{CHUANFENGLI, ANDREA}. Moreover, the information flow between the system and its environment and the corresponding degree of non-Markovianity is closely related to the presence of correlations \cite{WITNESS,NM,RIVAS,NMHEFEI}.

The present experiment follows a recently proposed protocol to detect nonclassical system-environment correlations of an arbitrary, unknown state by only accessing the open system \cite{GB,GB1}. The correlations are revealed through their effect on the open system dynamics. The protocol does not require any knowledge about the environment or the nature of the interaction, making it applicable to a wide range of scenarios where only partial access to a possibly correlated dynamical system is granted.

Previous experiments have detected correlations between system and environment in photonic systems \cite{CHUANFENGLI, ANDREA}, without distinguishing between classical and quantum correlations. In this work, we identify the detected correlations as quantum discord. A definition for this particular notion of quantum correlations will be given in the context of the local detection protocol. For pure total states, quantum discord is equivalent to entanglement. Quantum discord is considered a resource for certain quantum information processing protocols based on mixed states where little or no entanglement is needed \cite{MODIREVIEW,Lanyon,Dakic}.

\begin{figure}[tb]
\includegraphics[width=0.41\textwidth]{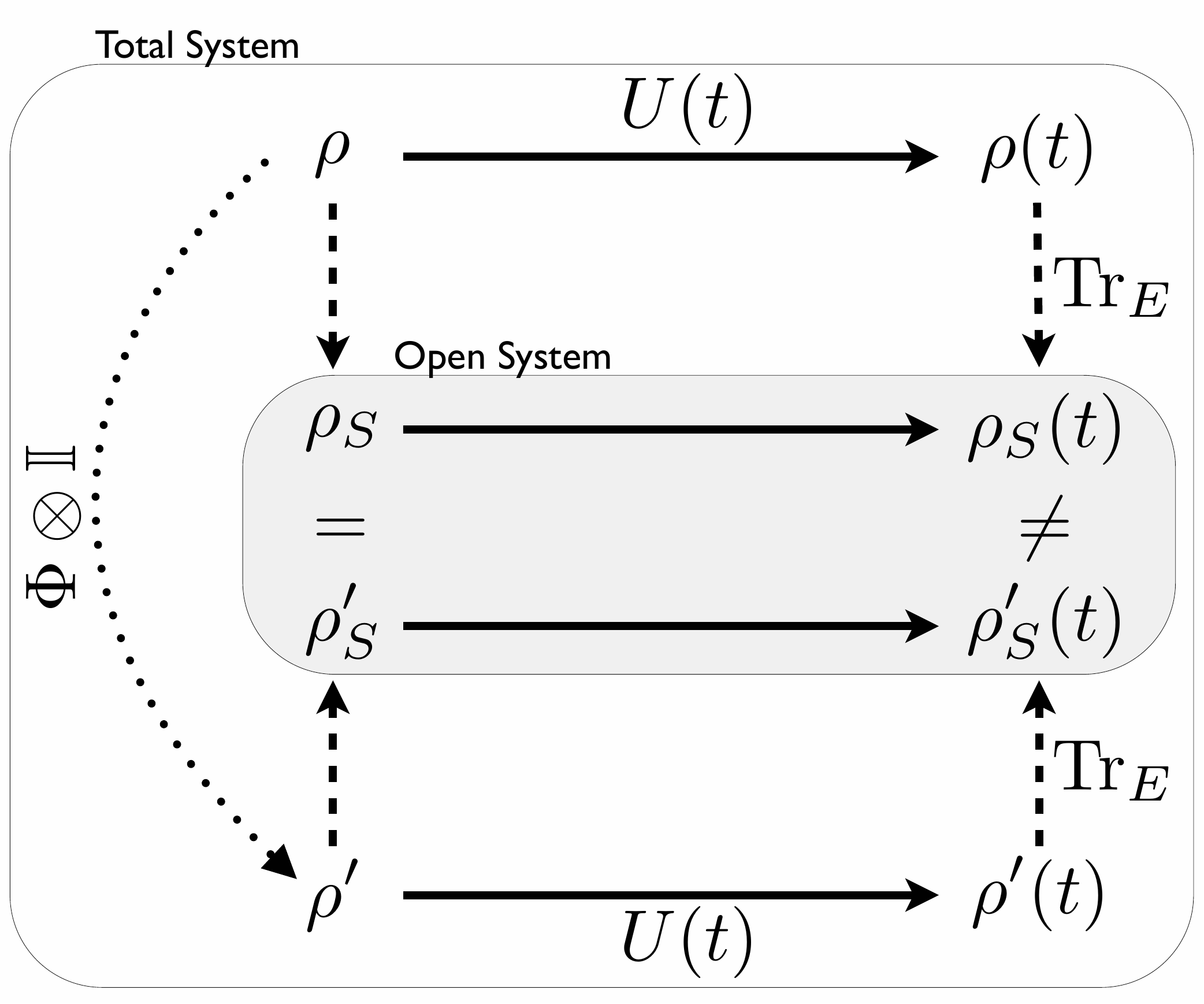}
\caption{\textbf{Outline of the local detection protocol.}
The outer box represents the total system, which is connected to its accessible local subsystem in the grey box via the partial trace operation $\mathrm{Tr}_E$ (dashed arrows).
First, the accessible part $\rho_S$ of the unknown state $\rho$ is measured. The eigenbasis of $\rho_S$ determines the local dephasing operation $\Phi\otimes\mathbb{I}$. Local dephasing removes the quantum correlations of $\rho$ and creates a reference state $\rho'$. The two states have the same initial reduced density matrices, $\rho_S=\rho'_S$. Now, both total states $\rho$ and $\rho'$ are subject to the same unitary time evolution $U(t)$. A different time evolution in the open system, i.e. $\rho_S(t)\neq\rho'_S(t)$, reveals the presence of quantum correlations in $\rho$. \cite{GB}}
\label{fig.theory}
\end{figure}

The local detection protocol is outlined in figure \ref{fig.theory}. It is based on the comparison of the time evolution of the locally accessible system with and without quantum correlations between the system and its environment. Any difference in these time evolutions proves the presence of quantum correlations. The first step consists in performing state tomography of the locally accessible part of the total state $\rho$, yielding the reduced density matrix $\rho_S=\mathrm{Tr}_E\rho$ of the system, where $\mathrm{Tr}_E$ denotes the partial trace over the environment \cite{BREUERBOOK}. On the basis of the eigenvectors $|i\rangle$ of $\rho_S=\sum_ip_i|i\rangle\langle i|$ we define the dephasing operation $\Phi$ as
\begin{align}\label{eq.dephasing}
\Phi(X)=\sum_i|i\rangle\langle i|X|i\rangle\langle i|.
\end{align}
This operation acts only on the accessible part of the state $\rho$, creating a reference state $\rho'=(\Phi\otimes\mathbb{I})\rho$, where $\mathbb{I}$ denotes the identity operation on the environment. The local dephasing operation $\Phi\otimes\mathbb{I}$ is the central element of the detection process: it is implemented on a strictly local level but erases all quantum correlations between the system and the environment. To see this, we first note that this operation does not change the reduced density matrices of either the system or the environment \cite{GB1}. The only difference between the states $\rho$ and $\rho'$ is the absence of certain coherences in $\rho'$. These coherences constitute the quantum correlations in $\rho$ according to the notion of quantum discord \cite{MODIREVIEW}. Hence, the state $\rho$ contains quantum discord if and only if $\rho\neq\rho'$. The next step of the protocol consists of subjecting both states $\rho$ and $\rho'$ to the same global unitary time evolution $U(t)$. We then compare how the subsystem evolves in time. More precisely, if the subsystem time evolution without quantum correlations,
\begin{align}
\rho'_S(t)=\mathrm{Tr}_E\left\{U(t)\rho' U^{\dagger}(t)\right\},
\end{align}
differs from the original time evolution,
\begin{align}\rho_S(t)=\mathrm{Tr}_E\left\{U(t)\rho U^{\dagger}(t)\right\},
\end{align}
one has detected non-vanishing quantum discord in the state $\rho$.

We apply the described protocol to a trapped ion system, consisting of an electronic two-level system coupled to a single mode of the ion's motion. The electronic state of the ion represents the open system, whereas we regard the ion motion as a simple environment. Using this model system allows us to establish quantum correlations in a well-controlled manner and, thus, to assess the performance of the protocol accurately.

\begin{figure}
\includegraphics[width=0.49\textwidth]{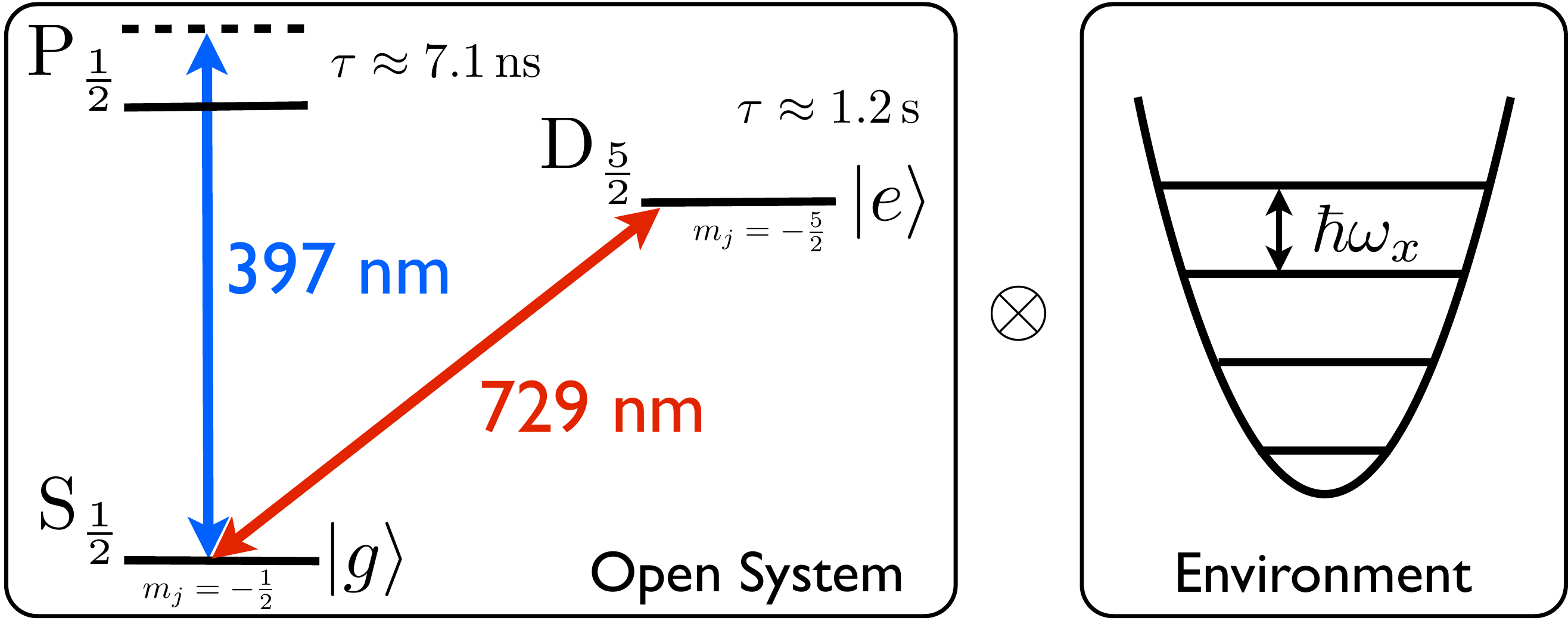}
\caption{\textbf{Description of open system and environment.} The left side depicts the relevant electronic levels of $^{40}\mathrm{Ca}^+$. We use a narrow linewidth 729\,nm laser on the quadrupole transition between the $|g\rangle=\left|S_{1/2},m_j=-\frac{1}{2}\right\rangle$ and $|e\rangle=\left|D_{5/2},-\frac{5}{2}\right\rangle$ states for qubit manipulations. A magnetic field of 100\,$\mathrm{\mu T}$ along the axial direction lifts the degeneracy of different Zeeman levels. The ion is optically pumped into $|g\rangle$. State readout and Doppler cooling are performed using the 397\,nm transition \cite{Leibfried}. An additional blue-detuned laser on the same transition implements the local dephasing operation via the AC-Stark effect. Detuning the 729\,nm laser to the blue sideband couples the electronic state to the ion's motion, described by a quantum harmonic oscillator.}
\label{fig.levelscheme}
\end{figure}

For our experiments, we trap a single $^{40}\mathrm{Ca}^+$ ion in a linear Paul trap. We encode a qubit in the two-level system consisting of the states $|g\rangle=\left|S_{1/2},m_j=-\frac{1}{2}\right\rangle$ and $|e\rangle=\left|D_{5/2},m_j=-\frac{5}{2}\right\rangle$, coherently manipulated with narrow band laser light at 729~nm, c.f. figure \ref{fig.levelscheme}. The frequencies $(\omega_x,\omega_y,\omega_z)$ of the harmonic ion motion are $2\pi\times(2.8,2.6,0.2)$\,MHz.

Correlations between the electronic state and the motion can be created by detuning the laser from the qubit transition to one of the motional sidebands. In particular, choosing a blue detuning of $+\omega_x$ generates the anti-Jaynes-Cummings Hamiltonian (see Supplementary Information) \cite{Leibfried}
\begin{align}
H\simeq\sum_n\left(\frac{\hbar\Omega_n}{2}\sigma_+|n+1\rangle\langle n|+\mathrm{H.c.}\right).
\end{align}
We have introduced the effective Rabi frequency $\Omega_n$,  eigenstates $|n\rangle$ of the harmonic oscillator, $\sigma_+=| e\rangle\langle g|$ and $\hbar=h/2\pi$, where $h$ is Planck's constant. This Hamiltonian couples the pairs of states $| g,n\rangle$ and $| e,n+1\rangle$.

\begin{figure}[t!]
\includegraphics[width=.43\textwidth]{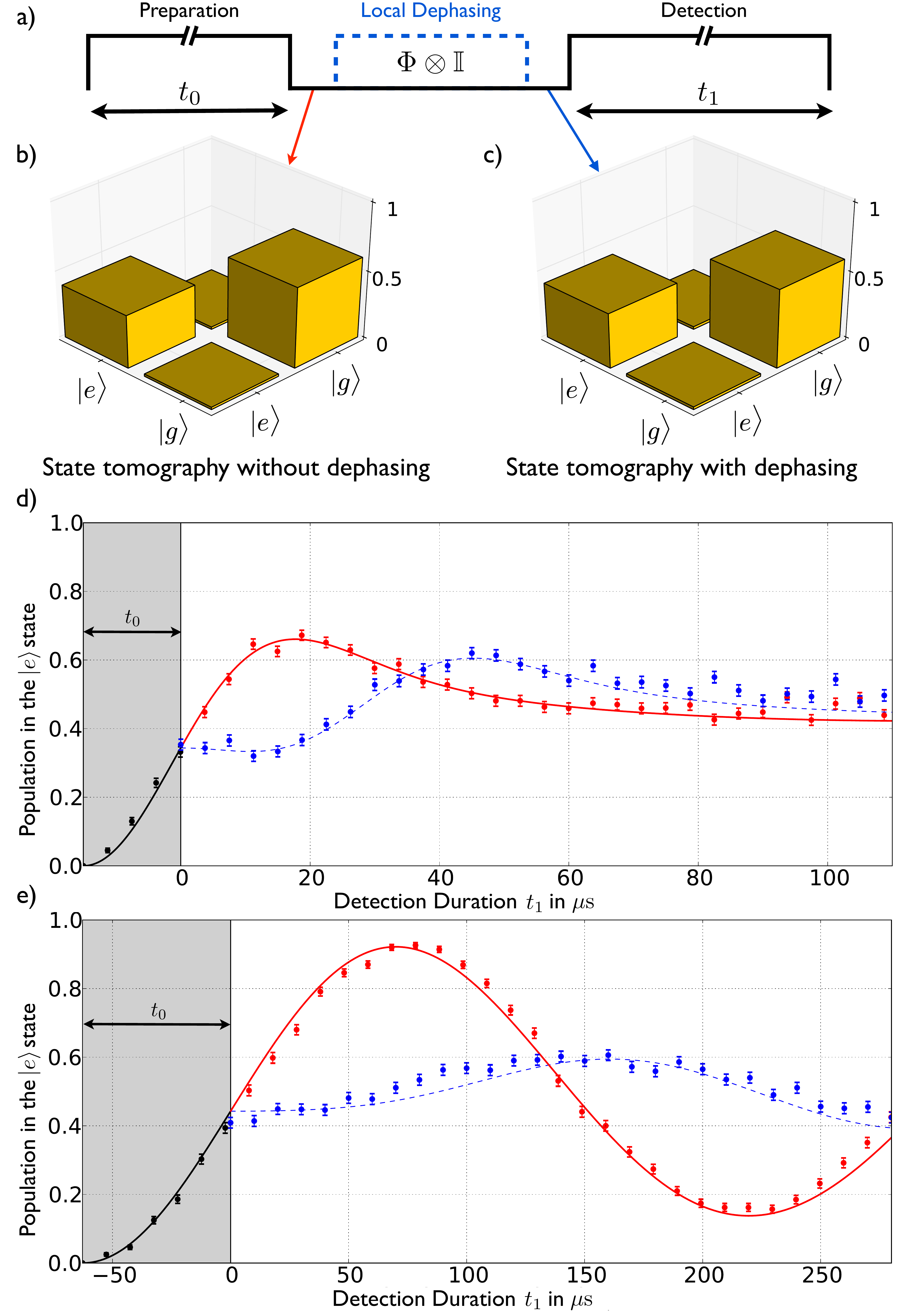}
\caption{\textbf{Local detection of qubit-motion quantum correlations.} The experimental sequence is illustrated in a). After a preparation sideband pulse of duration $t_0$, state tomography yields the reduced density matrix $\rho_S$ of the initial state (absolute values plotted in b). The evolution of the excited state population $\langle e|\rho_S(t)|e\rangle$ is observed under the subsequent sideband interaction of duration $t_1$ (plotted in red). This evolution is compared to a second realization where the local dephasing operation (\ref{eq.dephasing}) is applied in between the two pulses (blue). Comparisons of the time evolution of the excited state population are shown for a Doppler-cooled state ($\bar{n}=5.9$) (d) and a sideband-cooled state ($\bar{n}=0.2$) (e). Even though the initial reduced density matrix is the same with and without dephasing (see b and c), the time evolution is noticeably different. We fit $\langle e|\rho_S(t)|e\rangle$ to a theoretical model which is outlined in the Supplementary Information (red / black dots and lines). The obtained fit parameters determine the predicted evolution for the dephased state (dashed blue line) which is compared to the measured data (blue dots). The error bars display the statistical errors $\sigma_p=\sqrt{p(1-p)/n}$, where $n=1000$ is the number of measurements for each data point.}
\label{fig.piover2}
\end{figure}

\begin{figure*}[tb]
\includegraphics[width=0.9\textwidth]{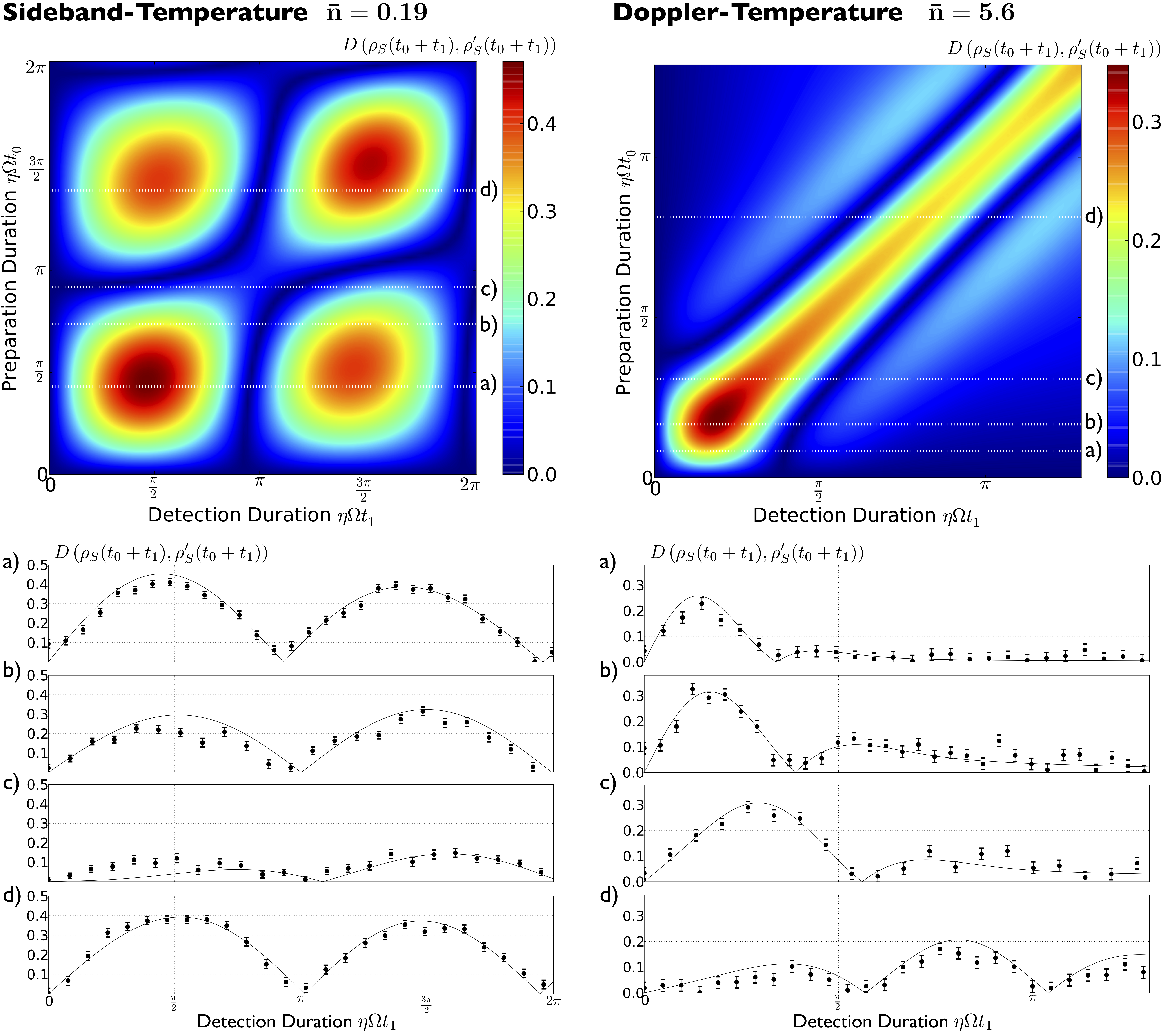}
\caption{\textbf{Open-System trace distance for environmental states of different temperatures (left side: $\bar{n}=0.19\pm0.02$, right side: $\bar{n}=5.6\pm0.5$).} The contour plots show the theoretical prediction for the trace distance of the open-system states, which serves as a witness for quantum correlations. The measured distance evolution after different preparation durations is shown in comparison to the theoretical curve in the subfigures. The time axis is scaled by the characteristic sideband interaction strength for the ground state $\eta \Omega$ with $\eta=0.04$ being the Lamb-Dicke factor and $\Omega \approx 2\pi\times100\,$kHz the Rabi frequency of the carrier. The black lines are the theoretical predictions for the parameters obtained by fitting $\langle e|\rho_S(t)|e\rangle$ for each of the preparation durations. The contour plot is generated with the average parameters of the individual measurements. In the sideband-cooled state, the time evolution is mostly determined by the ground state contribution. This makes the time evolution more periodic as opposed to the case of a Doppler-cooled state when multiple motional states with different Rabi frequencies contribute. The error bars report statistical errors of $1000$ runs for each point.}
\label{fig.2d}
\end{figure*}

Using a combination of Doppler and sideband cooling, we prepare thermal states of motion, resulting in the total state
\begin{align}
\rho_{0}=\sum_np_n|g,n\rangle\langle g,n|,
\end{align}
with $p_n=\bar{n}^n/(\bar{n}+1)^{n+1}$ and $\bar{n}$ denotes the mean occupation number of the motional state \cite{Leibfried}. A correlated state $\rho(t_0)$ is then created by driving the blue sideband transition for a time $t_0$. We first determine the density matrix of the local state, i.e. the qubit, $\rho_S=\mathrm{Tr}_E\rho(t_0)$, see figure \ref{fig.piover2}. As expected from theoretical considerations, we find that the eigenbasis of the qubit is given by the computational basis $\{|g\rangle,|e\rangle\}$ for all $t_0$ (c.f. Supplementary Information). Hence, local dephasing (\ref{eq.dephasing}) must be implemented in this basis. To this end, we shift the ground state energy by $h \times 40$~kHz with an AC-Stark shift generated by laser light detuned by +400\,GHz with respect to the S$_{1/2}$-P$_{1/2}$ transition. By varying the interaction time, we generate different phase shifts between $|e\rangle$ and $|g\rangle$. We sample over different phases between $0$ and $2\pi$ such that the phase factors average to zero, effectively removing all coherences in the basis $\{|g\rangle,|e\rangle\}$. We show in the Supplementary Information that this method can achieve dephasing in an arbitrary basis if it is combined with two unitary rotations.

Figures \ref{fig.piover2} b) and \ref{fig.piover2} c) show the density matrix before and after the dephasing and confirm that this operation leaves the local state unaffected. The estimated scattering rate is less than $10^{-3}$ photons during the dephasing and, thus, also the motional state remains unaltered (see Supplementary Information for details).

In order to detect the presence of quantum correlations, we compare the dynamics of the qubit with and without dephasing. Figures \ref{fig.piover2} d) and e) compare the time evolution of the excited state population under blue sideband interaction for different temperatures of the environment over a time interval $t_1$. We find pronounced differences in the open-system dynamics of the qubit, demonstrating the presence of quantum discord in the initial state. The signature of the correlations is more dominant when the ion was prepared in a nearly pure state by sideband-cooling (see figure \ref{fig.piover2} e). If the total system is known to be in a pure state, quantum correlations in form of entanglement can be detected by observation of mixed states in the open system \cite{HHHH}. However, the presented method does not require any assumptions about the total state and works well even for thermal states of higher temperature, as shown in figure \ref{fig.piover2} d). The evolution is in good agreement with the theoretical prediction for both temperature regimes.

\begin{figure}[tb]
\includegraphics[width=0.49\textwidth]{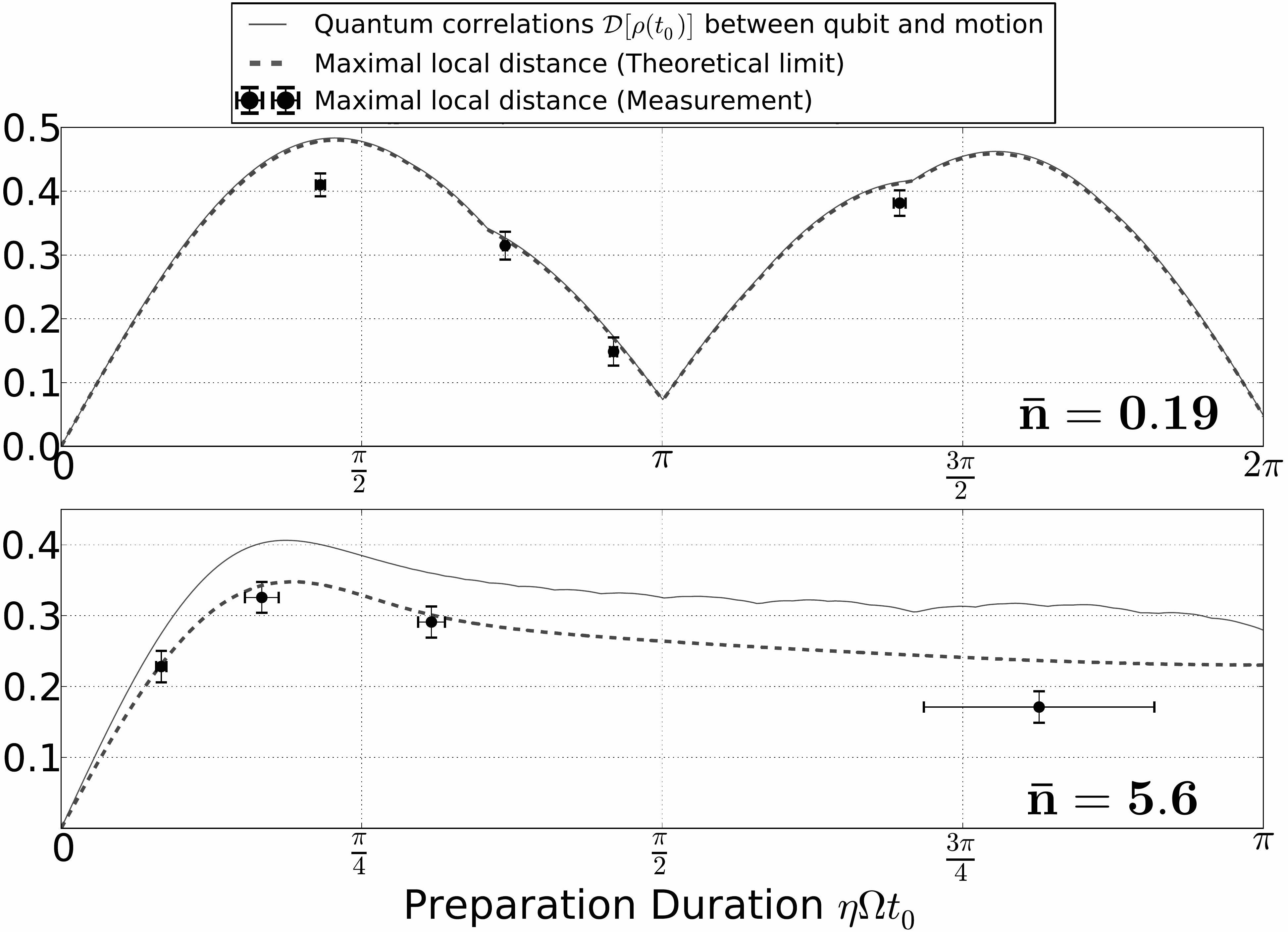}
\caption{\textbf{Maximum of the local distance (black) and quantum correlations at the dephasing time (blue).} The maximum of the local distance provides a lower bound to the amount of quantum correlations present at the time the dephasing was employed, see equation~(\ref{eq.maxdist}). For a state with the average properties of the four sideband-cooled realizations ($\bar{n}=0.19\pm0.02$) we find the lower bound almost reaching the actual value of the quantum correlations. Even for the average Doppler-cooled state ($\bar{n}=5.6\pm0.5$), the experimentally obtained maximal distance provides a reasonably tight lower bound for the quantum correlations and in most cases reaches the theoretical limit, which is obtained by numerical simulations. For the low-temperature state the theoretical maximum is not attained within the measured range for the detection time, which is displayed in figure \ref{fig.2d}. The error bars contain statistical errors (y-axis) and the error for the fit of the Rabi frequency (x-axis).}
\label{fig.maxdist}
\end{figure}
We measure the signature of the correlations on the reduced dynamics as a function of preparation duration $t_0$ and detection duration $t_1$, see figure \ref{fig.2d}. We quantify the difference in the time evolution by means of the trace distance $D(\rho_1,\rho_2)=\frac{1}{2}\|\rho_1-\rho_2\|$ of the reduced states,
\begin{align}\label{eq.trd}
D\left(\rho_S(t_0+t_1),\rho'_S(t_0+t_1)\right) = |d(t_0,t_1)|,
\end{align}
where $\|X\|=\mathrm{Tr}\sqrt{X^{\dagger}X}$ denotes the trace norm and $d(t_0,t_1)=\langle e|\rho_S(t_0+t_1)-\rho'_S(t_0+t_1)|e\rangle$ the difference of populations in the excited state.

Since the only difference between the original state $\rho(t_0)$ and the dephased state $\rho'(t_0)$ is the lack of quantum discord in $\rho'(t_0)$, we can use
\begin{align}\label{eq.trdiscord}
\mathcal{D}\left[\rho(t_0)\right]=D\left(\rho(t_0),\rho'(t_0)\right)
\end{align}
to quantify the quantum correlations of $\rho(t_0)$. \cite{Luo} Due to the contractivity property of the trace distance \cite{NIELSEN}, the local distance (\ref{eq.trd}) provides, for every value of $t_1$, a lower bound for the quantum correlations $\mathcal{D}\left[\rho(t_0)\right]$. \cite{GB1} The best lower bound is found by maximizing the local distance over all values of $t_1$:
\begin{align}\label{eq.maxdist}
\max_{t_1}D\left(\rho_S(t_0+t_1),\rho'_S(t_0+t_1)\right)\leq \mathcal{D}\left[\rho(t_0)\right].
\end{align}
As displayed in figure \ref{fig.maxdist}, we find the experimentally obtained lower bound remarkably close to the actual quantum correlations for both environmental temperatures (see Supplementary Information for further details).

The techniques developed in this work can be broadly applied to unknown states of open systems interacting with an arbitrary environment. In this context, it is important to note that for strong system-environment couplings initial correlations are known to have a substantial influence on the open system dynamics even for large, realistic environments with an infinite number of degrees of freedom and a continuous spectral density \cite{BREUERBOOK,GRABERT}.

We envision to apply this method to characterize quantum phase transitions through observation of quantum correlations \cite{Fazio,Dillenschneider} in systems where full access to the quantum state is not feasible, for example large trapped-ion crystals \cite{Monroe} or cold atoms in optical lattices \cite{SINGLESPIN} (see Supplementary Information for further details). In addition, we believe that the demonstrated scheme will be helpful to experimentally identify situations where the standard master equation treatment will fail to describe an open system in contact with an inaccessible or even unknown environment.

\textbf{Acknowledgments}

This work was supported by the NSF CAREER program grant \# PHY 0955650. M. G. thanks the German National Academic Foundation for support. M. R. was supported by an award from the Department of Energy Office of Science Graduate Fellowship Program administered by ORISE-ORAU under Contract No. DE-AC05-06OR23100. A. B. acknowledges financial support by DFG and under the EU-COST action "Fundamental Problems in Quantum Physics".

\textbf{Author Contributions}

M. G., A. B., H.-P. B. and H. H. devised the experiment. M. G., M. R., T. P. and H. H. performed the experiment. M. G. analyzed the data. All authors contributed to discussion of results and manuscript preparation.

\clearpage
\section{Supplementary Information}
\subsection{Laser-induced qubit-motion interaction}
We consider a single ion in a harmonic trap potential subject to laser light on an effective two-level transition. The experimental setup is described in reference \cite{sLifetime}. The Hamiltonian describing the system is given by
\begin{align}
H(t)=H_0+H_I(t),
\end{align}
with $H_0=\hbar\omega\sigma_+\sigma_-+\hbar\omega_x a^{\dagger}a$ and
\begin{align}
H_I(t)=\frac{\hbar\Omega}{2}(\sigma_++\sigma_-)(e^{i(\vec{k}\cdot\vec{x}-\omega_L t)}+e^{-i(\vec{k}\cdot\vec{x}-\omega_L t)}).
\end{align}
The frequency of the ionic two-level transition $|g\rangle\leftrightarrow|e\rangle$ is denoted by $\omega$, the trap frequency is $\omega_x$, and the laser is tuned to $\omega_L=\omega+\Delta+\delta$, where $\omega\gg\Delta\gg\delta$. We use $\Delta$ to tune resonantly to the sideband transition. The small detuning $\delta$ accounts for experimental imperfections. We express the ion's position in terms of raising and lowering operators $a$ and $a^{\dagger}$ of the harmonic potential as $\vec{k}\cdot\vec{x}=\eta(a+a^{\dagger})$, with the Lamb-Dicke parameter
\begin{align}
\eta=k\sqrt{\frac{\hbar}{2 m \omega_x}}\cos\theta,
\end{align}
$\theta$ describing the angle between the direction of laser propagation and the motional axis. Transformation to an appropriate interaction picture and application of the rotating wave approximation yields the Hamiltonian
\begin{align}\label{eq.hint}
H(t)=\frac{\hbar\Omega}{2}\left(\sigma_+e^{i\vec{k}\cdot\vec{x}(t)}e^{-i(\Delta+\delta) t}+\mathrm{H.c.}\right),
\end{align}
with $\vec{k}\cdot\vec{x}(t)=\eta(a e^{-i\omega_x t}+a^{\dagger} e^{i\omega_x t})$.
If the detuning $\Delta$ corresponds to multiples of the trap frequency $\omega_x$, that is, $\Delta=m\omega_x$, $m\in\mathbb{Z}$, the laser is on resonance with a sideband transition. In this case, terms oscillating with a frequency faster than $\delta$ can be ignored after a second application of the rotating wave approximation, c.f. references~\cite{sBLOCKLEY,sVOGEL,sWINELAND,sGARDINER,sLeibfried}.

\subsection{Dynamics of the first blue sideband}
For $\Delta=\omega_x$ we resonantly address the first blue sideband transition. The Hamiltonian can be approximated by
\begin{align}\label{eq.h}
H(t)=\sum_n\left(i\frac{\hbar\Omega_n}{2}e^{-i\delta t}\sigma_+|n+1\rangle\langle n|+\mathrm{H.c.}\right),
\end{align}
coupling the pairs of states $\{|g,n\rangle,|e,n+1\rangle\}$ with the effective Rabi frequency
\begin{align}
\Omega_n=\eta\sqrt{n+1}\Omega e^{-\eta^2/2}\sum_{k=0}^{n}\frac{(-\eta^2)^kn!}{k!(k+1)!(n-k)!}.
\end{align}
We denote the matrix elements of the unitary time evolution operator,
\begin{align}
U(t_1,t_0)=\mathcal{T}\exp\left(-\frac{i}{\hbar}\int_{t_0}^{t_1}H(\tau)d\tau\right),
\end{align}
as
\begin{align}
&\quad u^{n}_{gg}(t_1,t_0)\notag\\:\!&=\langle g,n|U(t_1,t_0)|g,n\rangle=\langle e,n+1|U(t_1,t_0)|e,n+1\rangle^*\notag\\&=\left[\cos\left(\frac{\tilde{\Omega}_n (t_1-t_0)}{2}\right)-\frac{i\delta}{\tilde{\Omega}_n}\sin\left(\frac{\tilde{\Omega}_n (t_1-t_0)}{2}\right)\right]\notag\\&\qquad\times e^{i\delta (t_1-t_0)/2},
\end{align}
and
\begin{align}
u^{n}_{eg}(t_1,t_0):\!&=\langle e,n+1|U(t_1,t_0)|g,n\rangle\notag\\
&=-\langle g,n|U(t_1,t_0)|e,n+1\rangle^*\notag\\
&=\frac{\Omega_n}{\tilde{\Omega}_n}\sin\left(\frac{\tilde{\Omega}_n (t_1-t_0)}{2}\right)e^{-i\delta (t_1+t_0)/2}.
\end{align}
We have introduced the generalized Rabi frequency $\tilde{\Omega}_n=\sqrt{\Omega_n^2+\delta^2}$. With this, the evolution of an initial thermal state
\begin{align}
\rho_0=\sum_{n=0}^{\infty}p_n|g,n\rangle\langle g,n|,
\end{align}
is expressed as
\begin{align}
\rho(t_0)&=U(t_0,0)\rho_0U^{\dagger}(t_0,0)\\&=\sum_{n=0}^{\infty}p_{n}\left[|u^{n}_{gg}(t_0,0)|^2|g,n\rangle\langle g,n|\right.\notag\\&\hspace{1cm}\left.+u^{n}_{eg}(t_0,0)^*u^{n}_{gg}(t_0,0)|g,n\rangle\langle e,n+1|\right.\notag\\&\hspace{1cm}\left.+u^{n}_{eg}(t_0,0)u^{n}_{gg}(t_0,0)^*|e,n+1\rangle\langle g,n|\right.\notag\\&\hspace{1cm}\left.+|u^{n}_{eg}(t_0,0)|^2|e,n+1\rangle\langle e,n+1|\right].\notag
\end{align}
The corresponding reduced state of the two-level system is given by
\begin{align}\label{eq.reducedevolution}
\rho_S(t_0)&=\mathrm{Tr}_E\rho(t_0)\\&=\sum_{n=0}^{\infty}p_{n}\left[|u
^n_{gg}(t_0,0)|^2|g\rangle\langle g|+|u
^n_{eg}(t_0,0)|^2|e\rangle\langle e|\right].\notag
\end{align}
This state is always diagonal in the computational basis $\{|g\rangle,|e\rangle\}$. In the experiment, we measure the excited state populatation $p_e(t)=\sum_{n=0}^{\infty}p_{n}|u^{n}_{eg}(t,0)|^2$. Due to magnetic field noise and rf-instabilities the fits in figures 3 and 4 contain a detuning $\delta$ which fluctuates with a variance $\sigma_{\delta}$ around a center $\delta_0$. Typically, $\delta_0$ and $\sigma_{\delta}$ are of the order of $\mathrm{2\pi\times1\,kHz}$.

\subsection{Detecting quantum correlations}
For the detection of quantum discord, our measure for quantum correlations, a reference state $\rho'(t_0)$ is created by local dephasing in the eigenbasis of the reduced state, which for sideband transitions is always given by $\{|g\rangle,|e\rangle\}$. We obtain
\begin{align}
\rho'(t_0)&=(\Phi\otimes\mathbb{I})\rho(t_0)\\&=\sum_{n=0}^{\infty}p_{n}\left[|u^{n}_{gg}(t_0,0)|^2|g,n\rangle\langle g,n|\right.\notag\\&\hspace{1cm}\left.+|u^{n}_{eg}(t_0,0)|^2|e,n+1\rangle\langle e,n+1|\right].\notag
\end{align}
The state $\rho'(t_0)$ differs from $\rho(t_0)$ only by lacking quantum discord. We use the trace norm $\|X\|=\mathrm{Tr}\sqrt{X^{\dagger}X}$, to quantify the quantum correlations in $\rho(t_0)$ as
\begin{align}\label{eq.trdqd}
\mathcal{D}\left[\rho(t_0)\right]&=D\left(\rho(t_0),\rho'(t_0)\right)\notag\\
&=\frac{1}{2}\|\rho(t_0)-\rho'(t_0)\|\notag\\
&=\sum_{n=0}^{\infty}p_{n}|u^{n}_{eg}(t_0,0)u^{n}_{gg}(t_0,0)|
\end{align}
After dephasing at time $t_0$ and a subsequent time evolution until $t=t_1+t_0$, we obtain the excited state population
\begin{align}
&\qquad\langle e|\rho'_S(t_1+t_0)|e\rangle\notag\\&=\langle e|\mathrm{Tr}_E\{U(t_1+t_0,t_0)\rho'(t_0)U^{\dagger}(t_1+t_0,t_0)\}|e\rangle\notag\\
&=\sum_{n=0}^{\infty}p_{n}\left[|u^{n}_{gg}(t_0,0)|^2|u^{n}_{eg}(t_1+t_0,t_0)|^2\right.\notag\\&\left.\hspace{1cm}+|u^{n}_{eg}(t_0,0)|^2|u^{n}_{gg}(t_1+t_0,t_0)|^2\right].
\end{align}
Henceforth we will refer to $t_0$ as the preparation time and label the duration $t_1$ of the second pulse the detection time. The difference in populations of the original state and the dephased state is given by
\begin{align}\label{eq.exstatediff}
&\qquad d(t_0,t_1)\\:\!&=\langle e|\rho_S(t_1+t_0)-\rho'_S(t_1+t_0)|e\rangle\notag\\&=\sum_{n=0}^{\infty}p_{n}\left[u_{eg}^{n}(t_0,0)^*u_{gg}
^n(t_0,0)\right.\notag\\&\left.\hspace{1.5cm}\times u_{eg}
^n(t_1+t_0,t_0)u_{gg}
^n(t_1+t_0,t_0)+\mathrm{c.c.}\right]\notag
\end{align}
Since the coherences are zero, the local trace distance is given by
\begin{align}\label{eq.localdistance}
D(\rho_S(t_0+t_1),\rho'_S(t_0+t_1))=|d(t_0,t_1)|.
\end{align}

\subsection{Quantification of quantum correlations}
\subsubsection{Trace distance: Lower bound by maximum local distance}
The trace distance is contractive under positive and trace-preserving maps $\mathcal{E}$:
\begin{align}
D\left(\mathcal{E}\left[\rho(t_0)\right],\mathcal{E}\left[\rho'(t_0)\right]\right)\leq D\left(\rho(t_0),\rho'(t_0)\right).
\end{align}
Since the time evolution from $t_0$ to $t_0+t_1$ and the partial trace operation are both described by positive maps, the local distance (\ref{eq.localdistance}) at $t_0+t_1$ provides a lower bound for the distance of the total states (\ref{eq.trdqd}) at $t_0$, which in turn quantifies the quantum correlations after the preparation time. Thus, we obtain
\begin{align}\label{eq.ineq}
D(\rho_S(t_0+t_1),\rho'_S(t_0+t_1))\leq \mathcal{D}\left[\rho(t_0)\right],
\end{align}
for all times $t_1$. Obviously, the best bound is found by maximizing the local signal over $t_1$.

In the following we assume $\delta=0$, making the Hamiltonian (\ref{eq.h}) time-independent. According to equation~(\ref{eq.localdistance}), the local distance yields
\begin{align}\label{eq.localdistnodelta}
&D(\rho_S(t_0+t_1),\rho'_S(t_0+t_1))\notag\\=\:&\frac{1}{2}\sum_{n=0}^{\infty}p_n|\sin(\Omega_nt_0)\sin(\Omega_nt_1)|,
\end{align}
while from equation~(\ref{eq.trdqd}), we obtain for the quantum correlations
\begin{align}
\mathcal{D}\left[\rho(t_0)\right]=\frac{1}{2}\sum_{n=0}^{\infty}p_n|\sin(\Omega_nt_0)|.
\end{align}
One can immediately see that, if the initial state of the environment is a Fock state $|n_0\rangle$, the inequality (\ref{eq.ineq}) is saturated for a detection time $t_1=\pi/(2\Omega_{n_0})$. The initial environmental state in the upper plot of figure 5 of the main manuscript is close to the absolute ground state $|0\rangle$, which explains why the lower bound is remarkably tight in this case.

If we assume $t_1=t_0$, equation~(\ref{eq.localdistnodelta}) becomes
\begin{align}
&D(\rho_S(2t_0),\rho'_S(2t_0))\notag\\=\:&\frac{1}{2}\sum_{n}p_{n}\sin^2(\Omega_nt_0)=\frac{1}{2}p_e(2t_0),
\end{align}
with the excited state probability $p_e(t)=\langle e|\rho_S(t)|e\rangle$ as given in equation~(\ref{eq.reducedevolution}). This means, if preparation time and detection time are chosen equal, the highest contrast between the excited state populations is achieved when $t_0=t_1=t_m/2$, where $t_m$ denotes the time when the Rabi flop reaches its global maximum. Note that the global maximum may be found at $t_0\neq t_1$. 

In figure 5 (main manuscript) the dashed line shows the theoretical limit of the maximum local distance, which was obtained by sampling over a large set of values for $t_1$. For the high-temperature state the value of $t_1$ which maximizes the local signal is found within the experimentally scanned range. For the low-temperature state the lower bound can be improved by scanning further values of $t_1$, beyond the interval which is displayed in figure 4 (main manuscript). This is due to the small number of frequencies contributing in equation (\ref{eq.localdistnodelta}) for this state. Experimentally it is challenging to measure long $t_1$ times because of decoherence effects.

A simple intuition for the effect of the dephasing on the dynamics can be gained with the example of the environmental ground state. In this case a $\pi$-pulse is realized for $t_m=\pi/\Omega_0$. Maximal contrast is thus expected for $t_0=t_1=\pi/(2\Omega_0)$, corresponding to two $\pi/2$-pulses with the dephasing applied in between. The first $\pi/2$ pulse creates the maximally entangled state $U(t_0,0)|g,0\rangle = (|g,0\rangle+|e,1\rangle)/\sqrt{2}$. A second $\pi/2$ pulse would bring this state all the way to the excited state $U(t_1,0)U(t_0,0)|g,0\rangle=|e,1\rangle$. Dephasing of the entangled state can be achieved by changing the relative phase of the coherent superposition randomly and averaging over all different contributions, effectively realizing an incoherent mixture. The outcome after application of the second $\pi/2$ pulse to each individual contribution depends strongly on this phase. For example, if we consider the extreme case where the phase has been flipped to a state $(|g,0\rangle-|e,1\rangle)/\sqrt{2}$, we end up in the ground state $|g,0\rangle$ after the second pulse. The average of all these contributions leads to a constant excited state probability at $p_e\equiv1/2$ for the dephased state, while the original state shows regular Rabi flops between zero and one. The maximal difference is thus found to be $1/2$.

\subsubsection{Hilbert-Schmidt distance: Quantification through time average}
Providing a lower bound on quantum correlations using the trace distance does not require any prior knowledge or assumptions about the environment or the specific interaction. It also does not rely on the fact that in the present experiments, we have a fixed interaction between the open system and the environment which is used for the preparation of the correlations and for their detection. In this section we present an additional result, which can be derived only if additional knowledge about the interaction is given. Specifically, assuming on-resonance anti-Jaynes-Cummings interaction allows us to quantify the initial quantum correlations in terms of a correlation measure based on the Hilbert-Schmidt distance \cite{sfootnote}.

For this purpose we quantify the quantum correlations in $\rho(t_0)$ as
\begin{align}
\mathcal{D}_{HS}\left[\rho(t_0)\right]&=\|\rho(t_0)-\rho'(t_0)\|_2^2\notag\\&=2\sum_{n=0}^{\infty}p_{n}^2|u^{n}_{eg}(t_0,0)|^2|u^{n}_{gg}(t_0,0)|^2.
\end{align}
where $\|X\|_2^2=\mathrm{Tr}X^{\dagger}X$ denotes the squared Hilbert-Schmidt norm. The local squared Hilbert-Schmidt distance is given by
\begin{align}\label{eq.localhsdistance}
\|\rho_S(t_0+t_1)-\rho'_S(t_0+t_1)\|_2^2=2|d(t_0,t_1)|^2.
\end{align}

\begin{figure}[tb]
\includegraphics[width=0.48\textwidth]{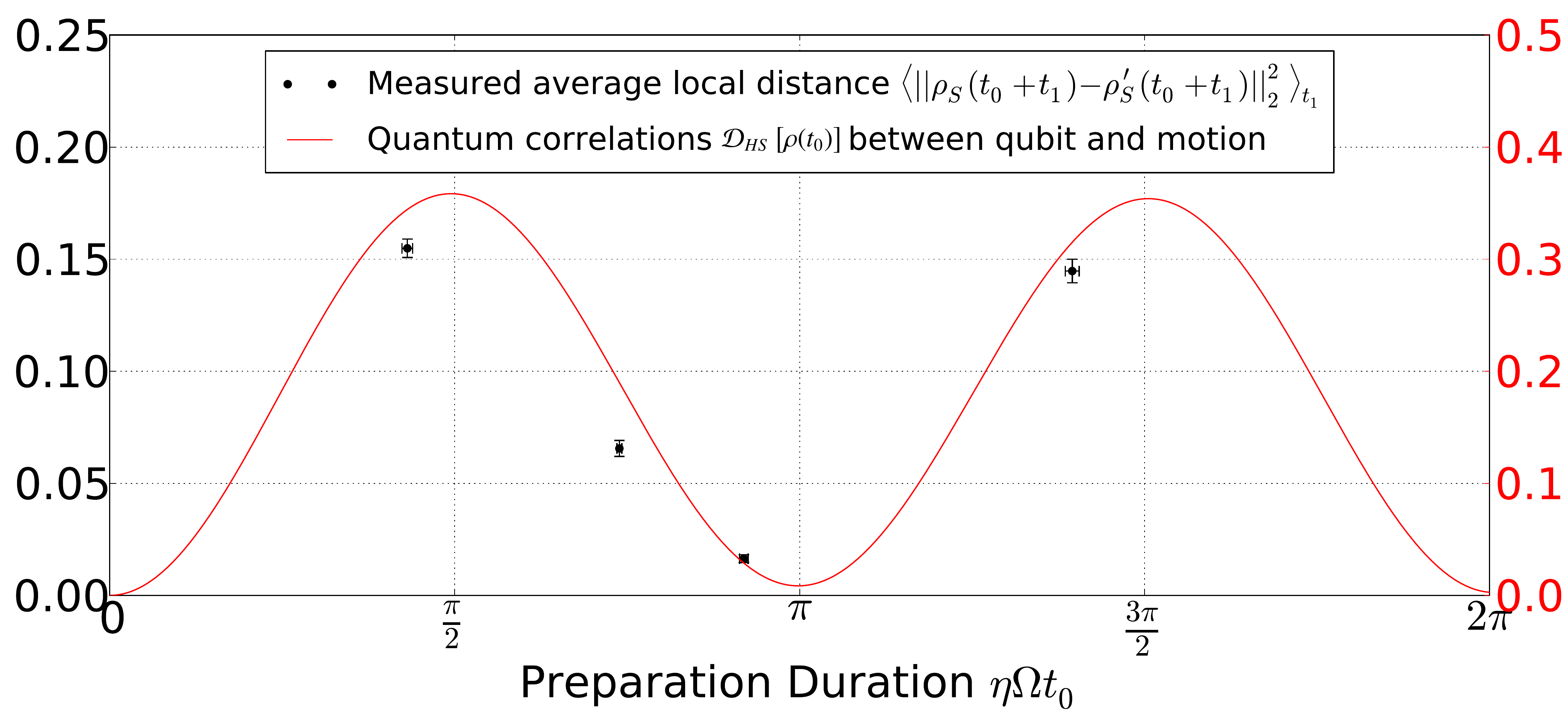}
\caption{\textbf{The time average of the local distance vs. Hilbert-Schmidt based quantum correlations at the dephasing time.} By averaging the local squared Hilbert-Schmidt distance, we quantify the amount of quantum correlations present at the time the dephasing was employed. For a state with the average properties of the four realizations ($\bar{n}=0.19\pm0.02$), we find the average distance to reflect the evolution of $\frac{1}{2}\mathcal{D}_{HS}\left[\rho(t_0)\right]$, c.f. equation (\ref{eq.average}). The measured contrast is a little lower than predicted due to laser beam pointing instabilities and fluctuating stray magnetic fields, causing additional decoherence.}
\label{fig.timeaverage}
\end{figure}
It has been shown that the average over all unitary time evolutions of the local signal (\ref{eq.localhsdistance}) is proportional to $\mathcal{D}_{HS}\left[\rho(t_0)\right]$ \cite{sGB,sGB1,sGB2}. Thus, the unitary average value could be used to quantify quantum correlations. Measuring the unitary average value is a very difficult task, however, the time average can be readily obtained. As in the ensemble-theoretic approach to statistical physics, there is assumed to be a proportionality between the time average and the unitary average \cite{sGB1}. Here we show that measuring the time-average over the detection time of the local system allows us to quantify the quantum correlations at the dephasing time. The open-system Hilbert-Schmidt distance reads
\begin{align}\label{eq.osdist}
&\qquad\|\rho_S(t_0+t_1)-\rho'_S(t_0+t_1)\|_2^2\\
&=\frac{1}{2}\sum_{n,m}p_{n}p_{m}\sin\left(\Omega_nt_0\right)\sin\left(\Omega_nt_1\right)\notag\\&\hspace{1.8cm}\times\sin\left(\Omega_mt_0\right)\sin\left(\Omega_mt_1\right).\notag
\end{align}
The time average over $t_1$ can be decomposed into terms of the form
\begin{align}
&\quad\left\langle\sin\left(\Omega_nt_1\right)\sin\left(\Omega_mt_1\right)\right\rangle_{t_1}\notag\\
&=\frac{1}{2}\lim_{T\rightarrow\infty}\frac{1}{T}\int\limits_0^Tdt_1[\cos((\Omega_n-\Omega_m)t_1)\notag\\&\hspace{2.5cm}-\cos((\Omega_n+\Omega_m)t_1)]\notag\\
&=\frac{1}{2}\delta_{\Omega_n\Omega_m}.
\end{align}
For frequencies which are relevant for low-temperature thermal states, $\Omega_n$ and $\Omega_m$ coincide only if $n=m$. Inserting this into equation~(\ref{eq.osdist}) yields
\begin{align}\label{eq.average}
&\left\langle\|\rho_S(t_0+t_1)-\rho'_S(t_0+t_1)\|_2^2\right\rangle_{t_1}\notag\\=\:&\sum_{n}p_{n}^2\cos^2\left(\frac{\Omega_nt_0}{2}\right)\sin^2\left(\frac{\Omega_nt_0}{2}\right)\notag\\=\:&\frac{1}{2}\|\rho(t_0)-\rho'(t_0)\|_2^2\notag\\=\:&\frac{1}{2}\mathcal{D}_{HS}\left[\rho(t_0)\right].
\end{align}
Hence, averaging the local distance over the detection time $t_1$ allows us to measure $\mathcal{D}_{HS}\left[\rho(t_0)\right]$.

\subsection{Implementing dephasing by far detuned laser light inducing an AC-Stark shift}
In order to locally remove quantum correlations from the initial state, we need to employ the operation
\begin{align}
\rho'=(\Phi\otimes\mathbb{I})\rho=\sum_{i\in\{e,g\}}(|i\rangle\langle i|\otimes\mathbb{I})\rho(|i\rangle\langle i|\otimes\mathbb{I}),
\end{align}
where $\{|g\rangle,|e\rangle\}$ are the eigenvectors of the reduced denstity matrix $\rho_S(t_0)$ at the dephasing time. This operation corresponds to a non-selective measurement in this basis, which is equivalent to full dephasing in this basis. Without controlling the motional state of the ion, this can be achieved with a far-detuned laser, inducing an AC-Stark shift on the ground state. In the limit of large detunings $\Delta$ the effective Hamiltonian, describing this Stark shift is given by \cite{sSchleich,sGerryKnight,sFoot}
\begin{align}
H_{\text{eff}}=\hbar\frac{\Omega^2}{4\Delta}(\sigma_+\sigma_-+\sigma_za^{\dagger}a).
\end{align}
The populations are not affected by this Hamiltonian, only the coherences will oscillate with angular frequency $\Omega^2/4\Delta$. Averaging over different interaction times (pulse lengths) removes the coherences without altering the populations.

We remark here that an equivalent dephasing effect could have been achieved by scanning the laser phase of the blue sideband laser. However, this laser is part of the interaction Hamiltonian, which in general will not be under experimental control. We therefore demonstrate a dephasing technique which can be implemented with purely local control on the open system.

\subsubsection{Experimental considerations}
When applying the above described sequence of detuned laser pulses in order to dephase the atomic levels, we need to tune the parameters of the Rabi frequency $\Omega$ and the detuning $\Delta$. We are limited by the constraint that at least one entire period of the oscillation must fit within the duration $t_{\text{max}}$ of the longest pulse. By averaging over pulses with different lengths between $t=0$ and $t=t_{\text{max}}$ a dephasing effect can be achieved. On the other hand, changes in the motional state due to scattering must be suppressed \cite{sOzeri,sUys}. The scattering rate is given by $\Gamma\rho_{ee}$, where
\begin{align}
\rho_{ee}=\frac{s/2}{1+s+(2\Delta/\Gamma)^2}
\end{align}
is the population in the $P_{1/2}$ state, $s=2\Omega^2/\Gamma^2$ is the saturation parameter and $\Gamma=1/\tau$ denotes the decay rate \cite{sLeibfried}. Since the scattering rate diminishes with higher detuning, it is favorable to detune as much as possible such that the oscillation period,
\begin{align}
T=\frac{8\pi\Delta}{\Omega^2},
\end{align}
still fits within the maximal pulse length $t_{\text{max}}$. The boundary condition $T=t_{\text{max}}$ leads to an optimal detuning of $\Delta=t_{\text{max}} \Omega^2/8\pi$ and a scattering rate of
\begin{align}\label{eq.scattering}
\Gamma\rho_{ee}=\frac{\Gamma s/2}{1+s+(t_{\text{max}}\Gamma s/8\pi)^2},
\end{align}
which decays with increasing saturation parameter $s$. The parameter $s$ can be expressed as $s=I/I_{\text{sat}}$, where \cite{sFoot}
\begin{align}
I_{\text{sat}}=\frac{\pi}{3}\frac{hc}{\lambda^3\tau}.
\end{align}
For the $\lambda=397\,\text{nm}$ transition of $^{40}\mathrm{Ca}^+$, we have $\tau\approx7.1\,\text{ns}$, which yields $I_{\text{sat}}=46.8\,\text{mW/cm}^2$. In the present experiment we have $t_{\text{max}}= 25\,\mu s$. For a detuning of $\Delta=2\pi\times400\,\text{GHz}$, this reflects the period of the above oscillation if $s\approx 255$. According to equation~(\ref{eq.scattering}), we expect an average of $3.5\times 10^{-4}$ scattering events during the maximal dephasing pulse length $t_{\text{max}}$.

\subsection{Applying dephasing in a different basis}
With the above method, dephasing is easily implemented in the $\{|g\rangle,|e\rangle\}$ basis. The dephasing can be performed equivalently in an arbitrary basis by additionally using unitary operations to rotate the basis.
Let us consider two bases $\{|\varphi_i\rangle\}$ and $\{|i\rangle\}$ of the open system Hilbert space, connected via the unitary transformation $U=\sum_i|\varphi_i\rangle\langle i|$. We assume that dephasing in the basis $|i\rangle$ is easily implemented and the unitary operation $U$ can be applied. In the case when dephasing is supposed to be implemented in the basis $|\varphi_i\rangle$, one can combine the dephasing in $|i\rangle$ with unitary rotations to achieve dephasing in $|\varphi_i\rangle$. More precisely, the intended dephasing operation
\begin{align}
\Phi_1(X)=\sum_i|\varphi_i\rangle\langle\varphi_i|X|\varphi_i\rangle\langle\varphi_i|,
\end{align}
can be achieved by a combination of the dephasing
\begin{align}
\Phi_2(X)=\sum_i|i\rangle\langle i|X|i\rangle\langle i|,
\end{align}
and the transformation $U$:
\begin{align}
&\qquad U\Phi_2(U^{\dagger}XU)U^{\dagger}\notag\\&=\sum_{ijklm}|\varphi_i\rangle\langle i|k\rangle\langle k|l\rangle\langle\varphi_l|X|\varphi_m\rangle\langle m|k\rangle\langle k|j\rangle\langle\varphi_j|\notag\\
&=\sum_{i}|\varphi_i\rangle\langle\varphi_i|X|\varphi_i\rangle\langle\varphi_i|=\Phi_1(X).
\end{align}
This means, in order to dephase in the basis $|\varphi_i\rangle$, one can first rotate $\rho_S$ by application of $U^{\dagger}$, then dephase in the basis $\{|i\rangle\}$, and finally rotate back with the inverse transformation $U$.

\subsection{Application of the method to different systems}

\subsubsection{Trapped-ion simulated Ising model}
Chains of trapped ions can be used to simulate Ising models with tunable spin-spin interactions by off-resonant lasers exerting an optical dipole force \cite{sPorras,sMonroe1}. Upon scanning the relative strength of the effective $B$-Field, the system undergoes a quantum phase transition \cite{sMonroe1}. Recent experiments have reached a large number of up to 16 ions \cite{sMonroe2} for which a full state tomography becomes experimentally impossible. Also the measurement of entanglement witness operators becomes experimentally demanding for such a large system. Thus, the quantum correlations have not been explicitly verified in these experiments.

Bipartite quantum correlations between a single spin and the rest of the chain can be detected applying the method presented in this paper. Individual ions in a Paul trap are routinely addressed with focussed lasers, which can be used to dephase and measure any ion in the chain. In the system described in references~\cite{sMonroe1,sMonroe2} radio-frequency qubits are represented by hyperfine ground states of $^{171}$Yb$^+$ ions. A state-dependent AC-Stark shift on the qubit can be implemented by a laser on the 369\,nm Doppler-cooling transition with $\sigma^-$-polarization. 

Applying the local detection protocol to each ion individually and considering the remaining ions as an environment provides a scalable method to detect and quantify quantum correlations in this type of system. This study of quantum correlations between a single site and the rest of the system during the phase transition is particularly compelling  because of a known correspondence between the behavior of entanglement and the critical point in the transverse Ising model \cite{sNielsen}.

\subsubsection{Neutral atoms in an optical lattice}
In this section we discuss application of the presented method to detect quantum correlations between the spin of a single neutral atom and its environment, consisting of a large number of identical atoms on an optical lattice. In particular, we show how the local dephasing operation on a single spin can be implemented with the setup described in reference~\cite{sSINGLESPIN}. In the described experiment, Rubidium atoms can be addressed individually by a tightly focussed laser entering through a microscope objective. A two-level system is encoded in the hyperfine levels $|g\rangle=|F=1,m_F=-1\rangle$ and $|e\rangle=|F=2,m_F=-2\rangle$ and coherent operations can be carried out with the aid of microwave radiation. Application of $\sigma^-$-polarized light with the 787.55\,nm addressing beam induces a local AC-Stark shift onto the upper level $|e\rangle$ while the ground state $|g\rangle$ remains unaffected. In reference~\cite{sSINGLESPIN} the authors report a state-dependent light shift of approximately $2\pi\times 70\,$kHz, which would enable to dephase the qubit within timescales on the order of $\mu$s.

This shows that implementation of the local dephasing operation on the spin degree of freedom of a single neutral atom is already experimentally feasible. Theoretical proposals for the generation of spin-dependent dynamics in the Mott-insulator regime \cite{sDuanLukin} have also been realized in experiments \cite{sAnderlini}.


\begin{thebibliography}{xx}

\bibitem{HHHH} Horodecki, R., Horodecki, P., Horodecki, M. \& Horodecki, K., Quantum entanglement, \textit{Rev. Mod. Phys.} \textbf{81}, 865--942 (2009).

\bibitem{MODIREVIEW} Modi, K., Brodutch, A., Cable, H., Paterek, T. \& Vedral,V., The classical-quantum boundary for correlations: Discord and related measures, \textit{Rev. Mod. Phys.} \textbf{84}, 1655--1707 (2012).

\bibitem{NIELSEN} Nielsen, M. A. \& Chuang, I. L., {\textit{Quantum Computation and Quantum Information}} (Cambridge University Press, Cambridge, 2000).



\bibitem{AUCCAISE} Auccaise, R., Maziero, J., C\'eleri, L. C., Soares-Pinto, D. O., deAzevedo, E. R., Bonagamba, T. J., Sarthour, R. S., Oliveira, I. S. \& Serra, R. M., Experimentally Witnessing the Quantumness of Correlations, \textit{Phys. Rev. Lett.} \textbf{107}, 070501 (2011).


\bibitem{SILVA} Silva, I. A., Girolami, D., Auccaise, R., Sarthour, R. S., Oliveira, I. S., Bonagamba, T. J., deAzevedo, E. R., Soares-Pinto, D. O. \& Adesso, G., Measuring Bipartite Quantum Correlations of an Unknown State, \textit{Phys. Rev. Lett.} \textbf{110}, 140501 (2013).

\bibitem{BREUERBOOK} Breuer, H.-P. \& Petruccione, F., {\textit{The Theory of Open Quantum Systems}} (Oxford University Press, Oxford,  2007).

\bibitem{PECHUKAS} Pechukas, P., Reduced Dynamics Need Not Be Completely Positive, \textit{Phys. Rev. Lett.} \textbf{73}, 1060 (1994).

\bibitem{LINDBLAD} Lindblad, G., On the existence of quantum subdynamics, \textit{J. Phys. A} \textbf{29}, 4197 (1996).


\bibitem{WITNESS} Laine, E.-M., Piilo, J. \& Breuer, H.-P., Witness for initial system-environment correlations in open-system dynamics, \textit{Europhys. Lett.} {\textbf{92}}, 60010 (2010).

\bibitem{GB} Gessner, M. \& Breuer, H.-P., Detecting Nonclassical System-Environment Correlations by Local Operations, \textit{Phys. Rev. Lett.} \textbf{107}, 180402 (2011).

\bibitem{GB1} Gessner, M. \& Breuer, H.-P., Local witness for bipartite quantum discord, \textit{Phys. Rev. A} \textbf{87}, 042107 (2013).






\bibitem{MYATT} Myatt, C. J., King, B. E., Turchette, Q. A., Sackett, C. A., Kielpinski, D., Itano, W. M., Monroe, C. \& Wineland, D. J., Decoherence of quantum superpositions through coupling to engineered reservoirs, \textit{Nature} \textbf{403}, 269--273 (2000).

\bibitem{SCHINDLER} Schindler, P., M\"uller, M., Nigg, D., Barreiro, J. T., Martinez, E. A., Hennrich, M., Monz, T., Diehl, S., Zoller, P. \& Blatt, R., Quantum simulation of dynamical maps with trapped ions, \textit{Nature Phys.} \textbf{9}, 361--367 (2013).

\bibitem{CHRISTIAN} Blatt, R. \& Roos, C. F., Quantum simulations with trapped ions, \textit{Nature Phys.} \textbf{8}, 277--284 (2012).




\bibitem{Fazio} Osterloh, A., Amico, L., Falci, G. \& Fazio, R. Scaling of entanglement close to a quantum phase transition, \textit{Nature} \textbf{416}, 608--610 (2002).


\bibitem{Dillenschneider} Dillenschneider, R., Quantum discord and quantum phase transition in spin chains, \textit{Phys. Rev. B} \textbf{78}, 224413 (2008).

\bibitem{HARTMUTREVIEW} H\"affner, H., Roos, C. F. \& Blatt, R., Quantum computing with trapped ions, \textit{Phys. Rep.} \textbf{469}, 155--203 (2008).

\bibitem{ABUREVIEW} Mintert, F., de Carvalho, A. R. R., Kus, M. \& Buchleitner,  A., Measures and dynamics of entangled states, \textit{Phys. Rep.} \textbf{415}, 207--259 (2005).

\bibitem{SINGLESPIN} Weitenberg, C., Endres, M., Sherson, J. F., Cheneau, M., Schau\ss, P., Fukuhara, T., Bloch, I. \& Kuhr, S., Single-spin addressing in an atomic Mott insulator, \textit{Nature} \textbf{471}, 319--324 (2011).

\bibitem{CHUANFENGLI} Li, C. F., Tang, J.-S., Li, Y.-L. \& Guo, G.-C., Experimentally witnessing the initial correlation between an open quantum system and its environment, \textit{Phys. Rev. A} \textbf{83}, 064102 (2011).

\bibitem{ANDREA} Smirne, A., Brivio, D., Cialdi, S., Vacchini, B. \& Paris, M. G. A., Experimental investigation of initial system-environment correlations via trace-distance evolution, \textit{Phys. Rev. A} \textbf{84}, 032112 (2011).

\bibitem{NM} Breuer, H.-P., Laine, E.-M. \& Piilo, J., Measure for the Degree of Non-Markovian Behavior of Quantum Processes in Open Systems, \textit{Phys. Rev. Lett.} \textbf{103}, 210401 (2009).

\bibitem{RIVAS} Rivas, A., Huelga, S. F. \& Plenio, M. B., Entanglement and Non-Markovianity of Quantum Evolutions, \textit{Phys. Rev. Lett.} \textbf{105}, 050403 (2010).

\bibitem{NMHEFEI}  Liu, B.-H., Li, L., Huang, Y.-F., Li, C.-F., Guo, G.-C., Laine, E.-M., Breuer, H.-P. \& Piilo, J., Experimental control of the transition from Markovian to non-Markovian dynamics of open quantum systems, \textit{Nature Phys.} \textbf{7}, 931--934 (2011).

\bibitem{Lanyon} Lanyon, B. P.,  Barbieri, M., Almeida, M. P. \& White, A. G., Experimental Quantum Computing without Entanglement, \textit{Phys. Rev. Lett.} \textbf{101}, 200501 (2008).

\bibitem{Dakic} Daki\'c, B., Lipp, Y. O., Ma, X., Ringbauer, M., Kropatschek, S., Barz, S., Paterek, T., Vedral, V., Zeilinger, A., Brukner, \v{C}. \& Walther, P., Quantum discord as resource for remote state preparation, \textit{Nature Phys.} \textbf{8}, 666--670 (2012).

\bibitem{Leibfried} Leibfried, D., Blatt, R., Monroe, C. \& Wineland, D., Quantum dynamics of single trapped ions, \textit{Rev. Mod. Phys.} \textbf{75}, 281--324 (2003).












\bibitem{Luo} Luo, S., Using measurement-induced disturbance to characterize correlations as classical or quantum, \textit{Phys. Rev. A} \textbf{77}, 022301 (2008).

\bibitem{GRABERT} Grabert, H., Schramm, P. \&  Ingold, G.-L., Quantum Brownian Motion: The Functional Integral Approach, \textit{Phys. Rep.} \textbf{168}, 115--207 (1988).


\bibitem{Monroe} Islam, R., Senko, C., Campbell, W. C., Korenblit, S., Smith, J., Lee, A., Edwards, E. E., Kim, Wang, C.-C. J., Freericks, J. K. \& Monroe, C., Emergence and Frustration of Magnetism with Variable-Range Interactions in a Quantum Simulator, \textit{Science} \textbf{340}, 583--587 (2013).



\end{thebibliography}

\begin{thebibliography}{xx}

\bibitem{sLifetime} Ramm, M., Pruttivarasin, T., Kokish, M., Talukdar, I. \& H\"affner, H., Precision Measurement Method for Branching Fractions of Excited $P_{1/2}$ States Applied to $^{40}$Ca$^+$, \textit{Phys. Rev. Lett.} \textbf{111}, 023004 (2013).

\bibitem{sBLOCKLEY} Blockley, C. A., Walls, D. F. \& Risken, H., Quantum Collapses and Revivals in a Quantized Trap, \textit{Europhys. Lett.} \textbf{17}, 509 (1992).

\bibitem{sVOGEL} Vogel, W. \& de Matos Filho, R. L., Nonlinear Jaynes-Cummings dynamics of a trapped ion, \textit{Phys. Rev. A} \textbf{52}, 4214 (1995).

\bibitem{sWINELAND} Meekhof, D. M., Monroe, C., King, B. E., Itano, W. M. \& Wineland, D. J., Generation of Nonclassical Motional States of a Trapped Atom, \textit{Phys. Rev. Lett.} \textbf{76}, 1796 (1996).

\bibitem{sGARDINER} Gardiner, S. A., Cirac, J. I. \& Zoller, P., Nonclassical states and measurement of general motional observables of a trapped ion, \textit{Phys. Rev. A} \textbf{55}, 1683 (1997).

\bibitem{sLeibfried} Leibfried, D., Blatt, R., Monroe, C. \& Wineland, D., Quantum dynamics of single trapped ions, \textit{Rev. Mod. Phys.} \textbf{75}, 281 (2003).

\bibitem{sfootnote} Note that the absolute information content of the quantum discord quantified through the Hilbert-Schmidt distance may be limited \cite{sPiani}. The trace distance based measure does not suffer from these limitations \cite{sPaula}.

\bibitem{sPiani} Piani, M., Problem with geometric discord, \textit{Phys. Rev. A} \textbf{86}, 034101 (2012).

\bibitem{sPaula} Paula, F. M., de Oliveira, T. R. \& Sarandy, M. S., Geometric quantum discord through the Schatten 1-norm, \textit{Phys. Rev. A} \textbf{87}, 064101 (2013).

\bibitem{sGB} Gessner, M. \& Breuer, H.-P., Detecting Nonclassical System-Environment Correlations by Local Operations, \textit{Phys. Rev. Lett.} \textbf{107}, 180402 (2011).

\bibitem{sGB1} Gessner, M. \& Breuer, H.-P., Local witness for bipartite quantum discord, \textit{Phys. Rev. A} \textbf{87}, 042107 (2013).

\bibitem{sGB2} Gessner, M. \& Breuer, H.-P., Generic features of the dynamics of complex open quantum systems: Statistical approach based on averages over the unitary group, \textit{Phys. Rev. E} \textbf{87}, 042128 (2013).

\bibitem{sSchleich} Schneider, S., Herkommer, A. M., Leonhardt, U. \& Schleich, W., Cavity field tomography via atomic beam deflection, \textit{J. Mod. Opt} \textbf{44}, 2333 (1997).

\bibitem{sGerryKnight} Gerry, C. C. and Knight, P. L., \textit{Introductory Quantum Optics}, (Cambridge University Press, 2005).

\bibitem{sFoot} Foot, C. J., \textit{Atomic Physics}, (Oxford University Press, 2007).

\bibitem{sOzeri} Ozeri, R., Langer, C., Jost, J. D., DeMarco, B., Ben-Kish, A., Blakestad, B. R., Britton, J., Chiaverini,  J., Itano, W. M., Hume, D. B., Leibfried, D., Rosenband, T., Schmidt, P. O. \& Wineland, D. J., Hyperfine Coherence in the Presence of Spontaneous Photon Scattering, \textit{Phys. Rev. Lett.} \textbf{95}, 030403 (2005).

\bibitem{sUys} Uys, H., Biercuk, M. J., VanDevender, A. P., Ospelkaus, C., Meiser, D., Ozeri, R. \& Bollinger, J. J., Decoherence due to Elastic Rayleigh Scattering, \textit{Phys. Rev. Lett.} \textbf{105}, 200401 (2010).

\bibitem{sPorras} Porras, D. \& Cirac, J. I.,  Effective Quantum Spin Systems with Trapped Ions, \textit{Phys. Rev. Lett.} \textbf{92}, 207901 (2004).

\bibitem{sMonroe1} Islam, R., Edwards, E. E., Kim, K., Korenblit, S., Noh, C., Carmichael, H., Lin, G.-D., Duan, L.-M., Wang, C.-C. J., Freericks, J. K. \& Monroe, C., Onset of a quantum phase transition with a trapped ion quantum simulator, \textit{Nature Commun.} \textbf{2}, 377 (2011).

\bibitem{sMonroe2} Islam, R., Senko, C., Campbell, W. C., Korenblit, S., Smith, J., Lee, A., Edwards, E. E., Kim, Wang, C.-C. J., Freericks, J. K. \& Monroe, C., Emergence and Frustration of Magnetism with Variable-Range Interactions in a Quantum Simulator, \textit{Science} \textbf{340}, 583--587 (2013).

\bibitem{sNielsen} Osborne, T. J. \& Nielsen, M. A., Entanglement in a simple quantum phase transition, \textit{Phys. Rev. A} \textbf{66}, 032110 (2002).


\bibitem{sSINGLESPIN} Weitenberg, C., Endres, M., Sherson, J. F., Cheneau, M., Schau\ss, P., Fukuhara, T., Bloch, I. \& Kuhr, S., Single-spin addressing in an atomic Mott insulator, \textit{Nature} \textbf{471}, 319--324 (2011).

\bibitem{sDuanLukin} Duan, L.-M., Demler, E. \& Lukin, M. D., Controlling Spin Exchange Interactions of Ultracold Atoms in Optical Lattices, \textit{Phys. Rev. Lett.} \textbf{91}, 090402 (2003).

\bibitem{sAnderlini} Anderlini, M., Lee, P. J., Brown, B. L., Sebby-Strabley, J., Phillips, W. D. \& Porto, J. V., Controlled exchange interaction between pairs of neutral atoms in an optical lattice, \textit{Nature} \textbf{448}, 452--456 (2007).

\end{thebibliography}
\end{document}